\documentstyle[preprint,prl,aps]{revtex}
\begin{document}

\input epsf

\preprint{\vbox{\hbox {August 1999} \hbox{WM-99-114} } }

\title{\bf Fourth Generation $b^\prime$ decays into b + Higgs}

\author{\bf Marc Sher}
\address{Physics Department}
\address{College of William and Mary, Williamsburg VA  23187}
\maketitle
\begin{abstract}
If a fourth generation quark exists whose mass is below $255$ GeV, then the
only two-body charged current decay, $b^\prime\rightarrow c+W$, is doubly-Cabibbo
suppressed.  For this reason, CDF  has searched for the one-loop neutral current
decay $b^\prime\rightarrow b+Z$, assuming that the branching ratio into $b+Z$ is
$100\%$; an analysis giving the bounds on $m_{b^\prime}$ for smaller branching ratios is in preparation.  In this Brief Report, we examine the neutral current decay $b^\prime\rightarrow b+H$, 
which will occur if the Higgs mass is less than $m_{b^\prime}-m_b$.  Four different 
cases are examined: the sequential case, the non-chiral isosinglet case, the
non-chiral isodoublet case, and a two-Higgs model with flavor-changing neutral
currents.  In the first three of these, the rates for $b^\prime\rightarrow b+Z$ and
$b^\prime\rightarrow b+H$ are comparable, assuming comparable phase space factors;
in the fourth, $b^\prime\rightarrow b+H$ is completely dominant.  Thus, we
emphasize the importance of giving $b^\prime$ mass bounds as a function of the branching ratio into $b+Z$, since  the 
assumption of a $100\%$ branching ratio for $b^\prime\rightarrow b+Z$ may only
be valid if the Higgs mass is near or above the $b^\prime$ mass.
\end{abstract}
\newpage

The CDF  Collaboration\cite{CDF}   reported last year a search for the neutral current 
decay of the fourth generation $b^\prime$ quark into $b+Z$.  An earlier
result from the D0 
Collaboration\cite{D0} ruled out $b'$ masses up to $95$ GeV from $b^\prime
\rightarrow b+\gamma$; above that mass, $b^\prime\rightarrow b+Z$ dominates.
The CDF bound is
more stringent.  In Ref. \cite{CDF}, they  have excluded $b^\prime$ quarks with masses up to $148$ GeV, depending on the $b^\prime$ lifetime, and assuming that the branching ratio of $b^\prime\rightarrow b+Z$
 is $100\%$.  CDF has presented preliminary limits\cite{cdf2} on the $b^\prime$ production
cross-section times $(BR(b^\prime\rightarrow b+Z))^2$ as a function of the 
$b^\prime$ mass, and has excluded $b^\prime$ masses from $100$ GeV up to
$170$ GeV, for $BR(b^\prime \rightarrow b+Z)>71\%$.

 It might seem surprising that the neutral current decay, which 
occurs through one-loop in the standard model, could dominate the tree-level
charged current decay.  However the decay $b^\prime\rightarrow t+W$ is forbidden
for $b^\prime$ masses below $255$ GeV (and the three body phase space severely 
suppresses the decay into $t+W^*$ for $b^\prime$ masses below about $230$ GeV), and the decay
$b^\prime\rightarrow c+W$ is doubly-Cabibbo-suppressed.  If the mixing angle 
which connects across two generations is very small, which would not be
surprising,
then the decay $b^\prime\rightarrow b+Z$ could very well be dominant.  
If the $b^\prime$ quark is non-chiral, either an isosinglet or part of an isodoublet,
then the GIM violation will lead to a tree-level $b^\prime\rightarrow b+Z$ decay.
In that case, the neutral current decay will certainly be dominant.  A detailed
discussion of these possibilities, including a full set of formulae and plots, can be found in Ref. \cite{fhs}.   In the early work of Mukhopadhyaya and Roy
\cite{roy}, the neutral current decay of  sequential, isosinglet, isodoublet
and mirror quarks were considered.  Their bounds were below the $Z$ mass,
and thus the primary decay modes were into photons and virtual $Z$'s.

In this Brief Report, we look at a decay mode not considered in the above, 
$b^\prime\rightarrow b+H$, where $H$ is the Higgs boson.  This occurs in the
standard model at one-loop, if kinematically accessible, and in non-chiral
models at tree-level.  It is, of course, much more difficult to detect, since
the $H$ will decay into $b\overline{b}$, leading to a purely hadronic
signature (although see the discussion below).  However, it could suppress the overall $b^\prime\rightarrow b+Z$ branching
ratio, weakening the mass  bounds, even in the non-chiral case.
  We will look at four models:
the standard model with a sequential $b^\prime$ quark, a vectorlike isosinglet model,
a vectorlike isodoublet model, and a two-Higgs model with tree-level flavor
changing neutral currents.  The discussion of $b^\prime\rightarrow b+H$ in the
first two of these models is not new; extensive discussions have been 
previously published.  However these publications are ten years old, and we
now know the top mass, precision electroweak studies give constraints, and the
CKM angles are better known.  The decay in the latter two models has not
been discussed.

\subsubsection{Sequential Quarks}

The simplest realization of a fourth family is to add left-handed doublets
and right-handed singlets (with a right-handed neutrino necessary to give
the extra neutrino a large mass).  The first calculations of $b^\prime\rightarrow b+H$
 were carried out by Hou and Stuart\cite{hou1} and by Eilam, Haeri and Soni
\cite{ehs}.  A much more detailed analysis, which was the first to directly 
compare the rate with that of $b^\prime\rightarrow b+Z$, which made no assumptions about
mixing angles and which discussed the anomalous thresholds that occur in the
calculation, appeared in the subsequent work of Hou and Stuart\cite{hou2}.

First,  consider the ratio of the neutral current decay $b^\prime\rightarrow
b+Z$ to the charged current decay $b^\prime\rightarrow c+W$.   The former decay
depends on the mass of the $t^\prime$ quark and $|V_{tb^\prime}|$; the latter depends
on $|V_{cb^\prime}|$.  For a $t^\prime$ mass of $250$ GeV,
 the ratio is given by (see Ref \cite{fhs} for full expressions and a plot)
\begin{equation}
{\Gamma(b^\prime\rightarrow bZ)\over \Gamma(b^\prime\rightarrow cW)}= 0.005
{|V_{tb^\prime}|^2\over |V_{cb^\prime}|^2}
\end{equation}
For different $t^\prime$ masses, the ratio varies roughly as $(m_{t^\prime}^2-m_t^2)^2$
 (note the GIM cancellation when the masses are equal).  We thus see how
sensitive the ratio is to the mixing angles.  If one were to choose 
 $|V_{tb'}|/|V_{cb'}|$ to be the same as $|V_{cb}|/|V_{ub}|=13\pm 3$, then the
above ratio is between $0.5$ and $1.3$.  However, the large top quark mass might 
indicate a very large mixing angle between the third and fourth generations, 
leading to a much bigger ratio.  Thus,  the neutral current
$b^\prime\rightarrow b+Z$ decay is certainly similar to, and could dominate,
the charged-current decay. 

In the ratio of $b^\prime\rightarrow b+H$ to $b^\prime\rightarrow b+Z$, the mixing angles 
cancel, so there is less arbitrariness in the result.  The result, given by
Hou and Stuart\cite{hou2}, is a function of $M_H$, $m_t$, $m_{t^\prime}$ and $m_{b^\prime}$.  Hou and Stuart give plots of the 
partial widths as a function of $m_{b^\prime}$, 
for four different values of $M_H$, three different values of $m_t$ and two
different values of $m_{t^\prime}$.    Fortunately, one of the choices for $m_t$ was
$175$ GeV (the others were $75$ and $125$ GeV), and the dependence on $m_{t^\prime}$,
 while important for the individual rates, is very weak in the ratio.  For
$m_H=100$ GeV, the ratio of $b^\prime\rightarrow b+H$ to $b^\prime\rightarrow b+Z$ is
approximately $(1.0,1.4,1.7,2.0,2.5)$ for $m_{b^\prime}=(150,175,200,225,250)$, 
respectively.   For $m_H=150$ GeV, phase space suppression sets in, and the
ratio, for the same $b^\prime$ masses, is $(0,0.15,0.7,1.0,1.6)$, respectively.
One sees that the two rates are very similar.  For a Higgs mass of
$100$ GeV, and a sequential $b^\prime$ quark, the  assumption that the branching
ratio for $b^\prime\rightarrow b+Z$ is $100\%$ is not valid.  On the other hand,
for a Higgs mass of $150$ GeV or higher, it may be reasonable.

Could one improve upon Hou and Stuart's calculation?  We now know the top quark
mass (and can distinguish between the Yukawa coupling $\overline{MS}$ mass
and the pole mass), we know from precision electroweak data that the $t'$ mass
cannot be much bigger than the $b'$ mass,  we have a much better understanding
of the production cross sections for heavy quarks, and b-tagging in hadron 
colliders is much better understood.  

However, it would be premature to carry out this analysis.  The reason is that
a sequential fourth generation has virtually been ruled out by precision 
electroweak data.  Erler and Langacker\cite{erler} note that the $S$ parameter
 is in conflict with a degenerate fourth generation by over three standard
deviations, or $99.8\%$.  One can weaken this discrepancy slightly by 
making the fourth generation non-degenerate, but it appears very unlikely that
a sequential fourth generation can be accommodated, {\it if} it is the
only source of new physics.  One way around this discrepancy is to assume
that there is new physics which partially cancels the fourth generation 
contribution to the $S$ parameter (such as Majorana neutrinos, additional
Higgs doublets, etc.).  This certainly can be done, and thus searches for
a sequential fourth generation should continue.  However, this new physics
will likely also contribute to $b^\prime\rightarrow b+H$ and to $b^\prime\rightarrow b+Z$.
Thus, without some understanding of the new physics, carrying out a high
precision improvement of the Hou-Stuart analysis is premature.

\subsubsection{Non-chiral fermions}

Of much greater theoretical interest than a sequential fourth generation is
a non-chiral (isosinglet or isodoublet) fourth generation.  These happen
automatically in a wide variety of models, including $E_6$-unification
models, gauge-mediated supersymmetric models, the aspon CP-violation model
and so on.  The motivations for these non-chiral generations are discussed
in detail in Ref. \cite{fhs}.  They only contribute to the $S$ parameter at
higher order, and are thus completely in accord with precision electroweak 
studies.  Due to the GIM violation, these models have tree-level $b^\prime bH$ and
$b^\prime bZ$ vertices, and thus the charged-current decay of the $b'$ becomes
less competitive.  Without the $b^\prime\rightarrow b+H$ decay, the  assumption that
 the branching ratio of $b^\prime\rightarrow b+Z$ is $100\%$ would be completely
justified.  

Let us first consider the case in which $b^\prime$ is an isosinglet quark.  The
first discussion of $b^\prime\rightarrow b+H$ was given in 1989 by del Aguila, Kane and
Quiros (AKQ)\cite{akq}, who looked at the possibility of using this decay to 
detect a light Higgs (if the $b^\prime$ mass were less than $M_Z+m_b$, it would be
the primary decay mode).  This work was followed up by a more extensive analysis
by del Aguila, Ametller, Kane and Quiros (AAKQ)\cite{aakq}.   
A much later analysis
of the various phenomenological aspects of isosinglet quarks can be found in
the work of Barger, Berger and Phillips\cite{bbp}.

Following AKQ,  consider the case in which the $b^\prime$ only mixes with the
$b$.  The Higgs doublet gives the usual mass term $m_b\overline{b}_Lb_R+{\rm h.c.}$, 
as well as a term $m^\prime\overline{b}_Lb^\prime_R+{\rm h.c.}$.  In addition, there are
gauge invariant mass terms $M_{b^\prime}\overline{b^\prime}_Lb^\prime_R+{\rm h.c.}$ and
$M^\prime\overline{b^\prime}_Lb_R+{\rm h.c.}$.  The $2\times 2$ mass matrix can then
be diagonalized.  The resulting mixing then gives $b^\prime bZ$ and $b^\prime bH$ vertices,
which are proportional to $m^\prime$.  Thus, one gets tree level interactions, 
suppressed only by a single Cabibbo-type angle ($m^\prime/M_{b^\prime}$).   The angle cancels
in the ratio, giving
\begin{equation}
{\Gamma(b^\prime\rightarrow b+H)\over \Gamma(b^\prime\rightarrow b+Z)}=
{M_{b^\prime}^2\over M_{b^\prime}^2+2M^2_Z}\left({M_{b^\prime}^2-M_H^2\over M_{b^\prime}^2-M^2_Z}
\right)^2
\end{equation}
This ratio is unity in the limit of large $M_{b^\prime}$, and is $0.7$ times the
phase space factor for $M_{b^\prime}=200$ GeV.  There will also be a $b^\prime cW$ vertex
induced by mixing, but this will be doubly-Cabibbo suppressed, and thus should
be negligible.

One thus sees that, once again, the $b^\prime\rightarrow b+H$ decay is comparable to
the $b^\prime\rightarrow b+Z$, assuming the Higgs mass is not close to (or greater
than) the $b^\prime$ mass.   Again, the charged current $b^\prime\rightarrow c+W$ decay
is expected to be much smaller (and, as shown in Ref. \cite{fhs}, the  $b^\prime\rightarrow t+W^*$ decay will be negligible for all $b^\prime$ masses
below $300$ GeV).

Although the isosinglet case is theoretically preferred (since isosinglet
quarks automatically appear in all $E_6$ unified models, as well as all
models with a $5+\overline{5}$ of $SU(5)$), one can ask what happens
if the fourth generation quarks form an isodoublet. The ratio of $b^\prime\rightarrow b+H$ to
$b^\prime\rightarrow b+Z$ is the same as in the isosinglet case.  However, there is
one important difference.  Although the ratio is the same, the individual rates are
much smaller.  This is because the GIM mismatch in the isodoublet case occurs
in the right-handed sector, and there is a helicity suppression which suppresses the vertex by an additional factor of $m_b/M_{b^\prime}$.  This means that the
charged current decays become much more competitive.  
It is shown in Ref. \cite{fhs} that the three-body $b^\prime\rightarrow t+W^*$
decay becomes competitive with the $b^\prime\rightarrow b+Z$ decay for $b'$ masses
of $200$ GeV, and greatly exceeds it for masses above $220$ GeV.  For
lighter masses, the $b^\prime\rightarrow c+W$ decay will still be important, and
may dominate depending on the value of the $V_{cb^\prime}/V_{tb^\prime}$ ratio (as in
the sequential fermion case).

\subsubsection{Two-Higgs models}

In the standard two-Higgs doublet models, the so-called Model I or Model II,
the Yukawa couplings to a Higgs are multiplied by a factor of $\cos\alpha
\over \sin\beta$, where $\alpha$ is a Higgs mixing angle and $\beta$ is a
ratio of vacuum expectation values (depending on the specific model and the
specific fermion charge, $v_2/v_1$ will either be $\tan\beta$ or $\cot\beta$).
In most models, Higgs mixing is fairly small, so $\cos\alpha$ is near unity.
In all of the above  cases, this factor will change
 the ratio of $b^\prime\rightarrow b+H$ to $b^\prime\rightarrow
 b+Z$ by a factor which is of order one. (It can't enhance the Higgs decay
mode too much in the sequential case, since too large an enhancement will make the $b^\prime$ or $t^\prime$ Yukawa coupling non-perturbative.)

A bigger effect might be expected in  Model III\cite{chengsher}.  In this
model, unlike Models I and II, no discrete symmetry is imposed in order to
suppress tree level flavor-changing neutral currents (FCNC) and thus
FCNC arise, even in the sequential case.   The observed lack of large FCNC
in processes involving first-generation quarks is explained by noting that 
many models will have a FCNC coupling given by the geometric mean of the
Yukawa couplings of the two quarks.  In that case, the tree-level $b'bH$
coupling (neglecting Higgs mixing) is given by $g\sqrt{m_bm_{b'}}/\sqrt{2}M_W$.

How does this coupling affect the results?
In the isosinglet case, the Model III coupling is of the same order of magnitude as
the expected coupling induced by the GIM violation, and thus none of our
arguments change.  However, in the  sequential fermion case, the Model III coupling is
{\it much} larger than the one-loop induced $b^\prime bZ$ coupling.  Also, in the  isodoublet
case, the Model III coupling is much larger than the GIM-violation induced $b^\prime bZ$ coupling.  Thus, since the Higgs coupling is so much larger in these two models,  $b^\prime\rightarrow
 b+H$ will dominate all $b'$ decays.
We conclude that in Model III, with either a sequential or isodoublet $b'$, the $b'$ decay is dominated by $b^\prime\rightarrow b+H$, and thus  the CDF and D0  bounds are completely inapplicable.

\subsubsection{Conclusions}

Previous  searches for a fourth generation quark assume a $100\%$ branching 
ratio into $b+Z$.  The other neutral current decay,
$b^\prime\rightarrow b+H$, has been examined, in the sequential case, the isosinglet case, the isodoublet
 case and a two-Higgs model with tree-level FCNC.  In all of these cases,
the rate for $b^\prime\rightarrow b+H$ is comparable to, or greater than, 
$b^\prime\rightarrow b+Z$ if the Higgs is kinematically accessible.

Currently, the CDF collaboration\cite{cdf2} is preparing an analysis which 
will give the bounds as a function of the branching ratio to $b+Z$. 
This analysis is conservative in that it assumes that it is insensitive to
other decay channels than $b^\prime\rightarrow b+Z$.  However, suppose that  one $b^\prime$ decays to $b+Z$ and the other
to $b+H$.  At least one $b+Z$ decay is needed to trigger the event, and the
three $b$ final state of the other $b^\prime$ could then be detected. 
The $b$-tag efficiency in these events is expected to be considerably higher
than in $bbZZ$ events because of the 4-$b$ jets final state in which at
least two $b$-jets have high-$p_T$, independently of the $b'$ and Higgs
masses.
\cite{joao}.    This
leads to the potential exciting result that the experiment could discover
both a fourth-generation quark and a Higgs boson!

The only discouraging model is
 Model III, in the sequential or isodoublet cases.  Pair-production of $b^\prime$'s would lead to a 6-b final state, in
which every $b$ comes from a 2-body decay (except in the narrow region
of parameter space where $H\rightarrow W^+W^-$ can occur).  This would lead to quite dramatic
signatures, but without a lepton trigger, finding such a signature would be
very difficult.

I thank the CERN Theory Group for its hospitality while this work was
written, and Yao Yuan for her assistance.  I am also grateful to Jo\~{a}o
Guimar\~{a}es da Costa for informing me of the continuing interest of CDF in
conducting $b^\prime$ searches, and for many useful discussions.   This work was supported by the National Science Foundation.

\def\prd#1#2#3{{\rm Phys. ~Rev. ~}{D#1}, #3 (19#2)  }
\def\plb#1#2#3{{\rm Phys. ~Lett. ~}{B#1}, #3 (19#2)  }
\def\npb#1#2#3{{\rm Nucl. ~Phys. ~} {B#1}, #3 (19#2) }
\def\prl#1#2#3{{\rm Phys. ~Rev. ~Lett. ~}{#1}, #3 (19#2) }


\begin{thebibliography}{99} 
\bibitem{CDF}
F. Abe, et al., CDF Collaboration, \prd{58}{98}{051102}.
\bibitem{D0}
S. Abachi, et al., D0 Collaboration, \prl{78}{97}{3818}.
\bibitem{cdf2}
J. Guimar\~{a}es de Costa, Ph.D. thesis, University of Michigan, 1999, in
preparation; talk presented at the meeting of the American Physical
Society, Atlanta, March 1999. CDF Collaboration, in preparation, to be submitted to Physical Review Letters.
\bibitem{fhs}
P.H. Frampton, P.Q. Hung and M. Sher, Physics Reports, in press, 1999.
\bibitem{roy}
B. Mukhopadhyaya and D.P. Roy, \prd{48}{92}{2105}.
\bibitem{hou1}
W.S. Hou and R.G. Stuart, \plb{233}{89}{485}.
\bibitem{ehs}
G. Eilam, B. Haeri and A. Soni, \prd{41}{90}{875} and \prl{62}{89}{719}.
\bibitem{hou2}
W.S. Hou and R.G. Stuart, \prd{43}{91}{3669}.
\bibitem{erler}
J. Erler and P. Langacker, Eur. Phys. J. C3, 90 (1998).
\bibitem{akq}
F. del Aguila, G.L. Kane and M. Quiros, \prl{63}{89}{942}.
\bibitem{aakq}
F. del Aguila, Ll. Ametller, G.L. Kane and J. Vidal, \npb{334}{90}{1}.
\bibitem{bbp}
V. Barger, M.S. Berger and R.J.N. Phillips, \prd{52}{95}{1663}.
\bibitem{chengsher}
T.P. Cheng and M. Sher, \prd{35}{87}{3484}
\bibitem{joao}
J. Guimar\~{a}es de Costa, private communication.
\end{thebibliography}
\end{document}